\documentclass[
reprint,
twocolumn,
amsmath,amssymb,superscriptaddress,aps,
prl]{revtex4-1}

\usepackage{graphicx}
\usepackage{dcolumn}
\usepackage{bm}

\usepackage[utf8]{inputenc}
\usepackage[T1]{fontenc}
\usepackage{mathptmx}
\usepackage{etoolbox}
\usepackage{xcolor}
\usepackage{changes}
\usepackage{appendix}
\usepackage{tabularx}
\usepackage{multirow,hyperref}

\makeatletter
\def\@email#1#2{%
 \endgroup
 \patchcmd{\titleblock@produce}
  {\frontmatter@RRAPformat}
  {\frontmatter@RRAPformat{\produce@RRAP{*#1\href{mailto:#2}{#2}}}\frontmatter@RRAPformat}
  {}{}
}%
\makeatother
\begin{document}
\title{Node-weighted recurrence analysis for path dynamics on networks}

\author{A. Schmaus}
 \affiliation{Potsdam Institute for Climate Impact Research,Telegrafenberg A 31, Potsdam, Germany}
 \affiliation{Technical University Berlin, Straße des 17. Juni 135, Berlin 10623, Germany}

\author{N. Marwan}
\affiliation{Potsdam Institute for Climate Impact Research,Telegrafenberg A 31, Potsdam, Germany}

\author{N. Molkenthin}
\email{nora.molkenthin@pik-potsdam.de}
 \affiliation{Potsdam Institute for Climate Impact Research,Telegrafenberg A 31, Potsdam, Germany}

\date{\today}

\begin{abstract}
Trajectories of units moving on networks are relevant for nonlinear dynamical systems as diverse as polymers, ocean drifters, and human mobility. Although RQA is a well-researched tool with applications in many areas, it has rarely been used for spatial trajectories on networks. Here, we explore the use of RQA for paths on networks. We find that path dynamics on networks display recurrence patterns that are not often described in other applications of recurrence analysis. In particular, the combination of diagonal lines and perpendicular diagonal lines, indicates backtracking paths. 
We find that recurrence analysis for path dynamics on networks can be helpful to a) better understand the network structure if dynamic and recurrence plots are known, b) better understand the dynamics if network and recurrence plots are known, and c) understand the interaction between path dynamics and the underlying network.
\end{abstract}

\maketitle

\section{Introduction}

Trajectories in discrete or continuous space are the basis of interesting dynamics relevant to very diverse phenomena from polymer chains \cite{bhattacharjee2013flory,molkenthin2016scaling} to foraging behavior \cite{osborne2013ontogeny,picardi2020analysis} and from ocean drifters \cite{soreide2001overview,rypina2022applying} to human mobility \cite{helbing2001traffic, molkenthin2020scaling}. The models for such systems range from random walks \cite{codling2008random} to protein contact maps \cite{emerson2017protein}. In general, a trajectory is defined as the path of a moving object. A  trajectory combines a spatial and temporal aspect and could be described as a time series of spatial locations visited by the object.
An important analysis tool for time series and other complex systems is recurrence quantification analysis (RQA) \cite{marwan2007recurrence}, which has been applied to a wide range of time series data for
classifying dynamics, quantifying determinism and predictability, and detecting transitions \cite{marwan2023}. However, its applications to path trajectories on networks have been surprisingly rare. For polymer chains, which are conceptually similar but lack the temporal aspect, protein contact maps or protein residue networks \cite{webber2001,karain2017b,guven2023geometrically,lesne2025}, are mathematically equivalent to recurrence plots \cite{lesne2025} when interpreting the physical chain as the path of an object moving along the chain. 
In other words, where contact maps consider recurrences in space, path dynamics consider recurrences in time.
As a consequence, recurrence features, such as vertical lines, are strictly limited by spatial limitations.
In turn, contact maps may feature diagonal and anti-diagonal lines, with interpretations as protein secondary structure elements.

Here, we systematically explore RQA for trajectories on networks and gain a better understanding of the interaction of network structure and trajectory dynamics. The contributions of this paper are as follows:

\begin{enumerate}
    \item comprehensive recurrence quantification analysis of path dynamics on networks
    \item distinguishing between network and dynamics effects
    \item introduction of node-weighted tools for this analysis
\end{enumerate}

The trajectories are given as lists of nodes, which are visited in order. The trajectories are denoted as $T=\{u_{0}, u_{1}...,u_{f}\}$, where $u_i$ is the numerical identifier of the node visited at time step $i$. We denote the total length of the trajectory as $|T|$.
To account for the spatial aspect of the trajectories, the binary recurrence matrix is augmented with the node identifier value corresponding to the recurrence, similiar to chromatic, or symbolic recurrence plots \cite{cox2016chromatic, caballero2018symbolic}. We introduce the terms \emph{node-weighted recurrence matrix} and \emph{node-weighted recurrence plot}, to account for the spatial nature of the values:
\begin{equation}
    V_{i,j}=
    \begin{cases}
        u_i \text{     if  } u_i=u_j\\
        0 \text{       if  } u_i \neq u_j
    \end{cases}.
    \label{eq.valrec}
\end{equation}

\section{Definitions of dynamics and networks}
\begin{figure}[b]
    \centering
    \includegraphics[width=\columnwidth]{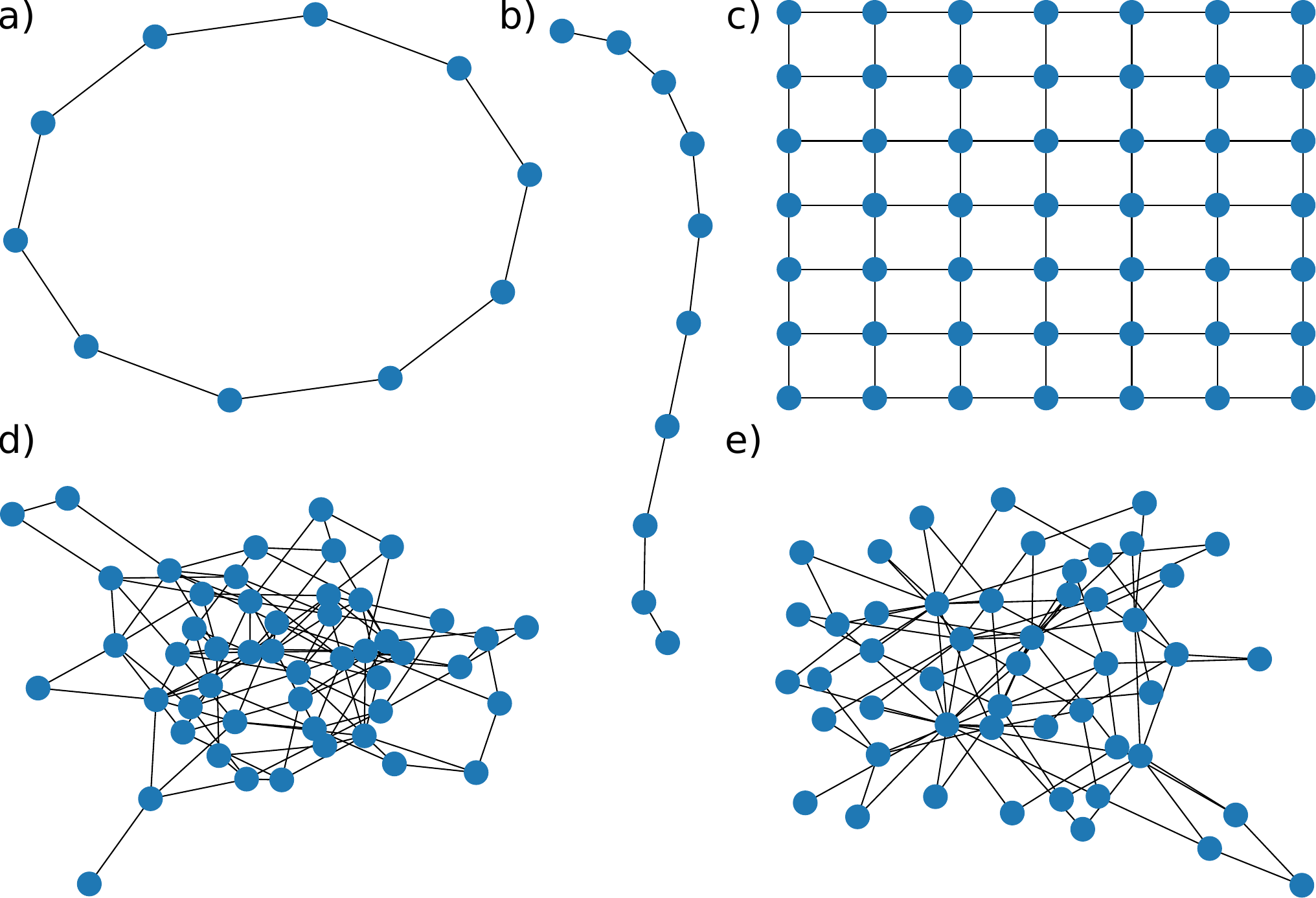}
    \caption{All path dynamics are evaluated on 5 network types: a) ring (10 nodes), b) line (10 nodes), c) grid (49 nodes), d) ER network (50 nodes), and e) BA network (50 nodes).}
    \label{fig:1}
\end{figure}
\begin{figure*}[htbp]
    \centering
    \includegraphics[width=\textwidth]{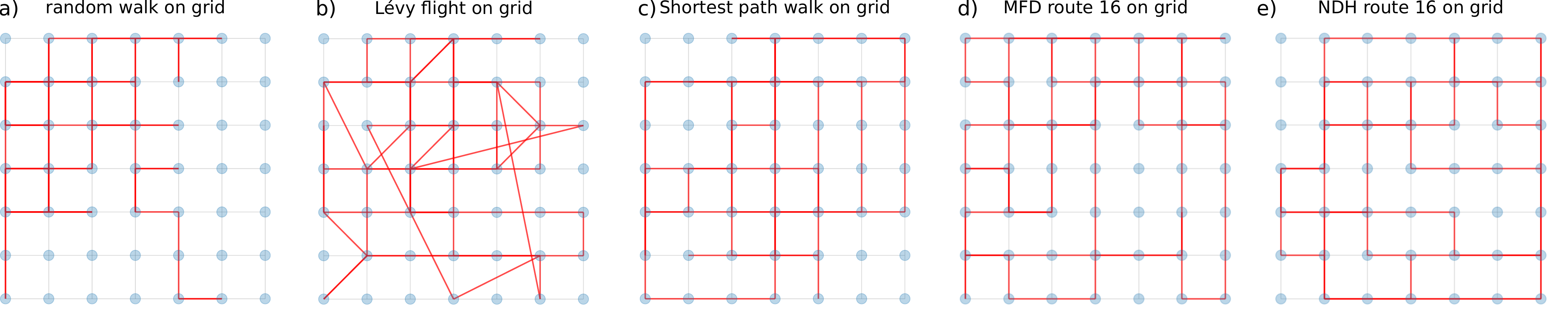}
    \caption{Example trajectories on $7\times 7$ grid for a) random walk, b) Levy flight, c) Shortest path walk, d) ride pooling trajectory with MFD dispatcher, e) ride pooling trajectory with NDH dispatcher. }
    \label{fig:2}
\end{figure*}
We have selected five network types to understand the impact of the underlying networks on the recurrence patterns of trajectories (Fig.~\ref{fig:1}): Ring, line, grid, Erdős-Renyi (ER), and Barabasi-Albert (BA) \cite{newman2003structure}. The ring and line networks are selected as examples for basic networks, on which many dynamics can be explored analytically. Ring and line networks offer limited options for trajectories. Thus, small networks of $N=10$ are sufficient to illustrate the full range of their recurrence patterns. 
The other three networks were set to sizes of $N\approx 50$, as a trade-off between being large enough to show sufficient differences in the interaction patterns but small enough to allow multiple realizations of all dynamics, fleet sizes, and parameters as well as covering all areas of the networks multiple times within the simulation time. The square grid (here $7\times 7$ nodes) is emblematic for networks embedded in 2D space. The ER and BA networks, each with 50 nodes, are used as examples of popular network models.
All networks are simple graphs without weights and directed links. The link densities are chosen so that the two smaller networks (a and b) and the three larger networks (c, d, and e) have approximately the same number of links. These networks do not cover all possible network types but span a broad spectrum of common examples, from 1D to scale-free.

We use the following path dynamics: a) simple unbiased random walk \cite{lovasz1995mixing}. b) Levy-flight, which is a random walk with a power-law probability for long-range jumps \cite{volchenkov2011random}. c) Shortest path random walk, which travels to randomly selected points along a shortest path, selecting a new target upon arrival. d) Minimum Fleet Distance dispatcher or MFD, which is constructed by simulating a fleet of paths, which are competing for uniformly random origin-destination pairs based on minimizing the distance added to the fleet. The OD pairs are generated as a Poisson process and inserted into one of the paths using a brute-force minimization of the added route length, while not exceeding a delay factor relative to the direct distance between origin and destination \cite{schmaus2025urban}.
e) No Detour Heuristic dispatcher or NDH, which is constructed by simulating a fleet of paths, which are competing for uniformly random origin-destination pairs based on minimizing the arrival time of the origin-destination pair. Here, insertions are only permitted if they do not insert distance before the route's last stop \cite{schmaus2022markov}. If no insertion is possible within a route, the new request is added to the end of the route. Among all paths, the one minimizing the request's arrival time is selected. An example trajectory for each dynamic on a square grid is given in Fig.~\ref{fig:2}.
\begin{figure*}[t]
    \centering
    \includegraphics[width=\textwidth]{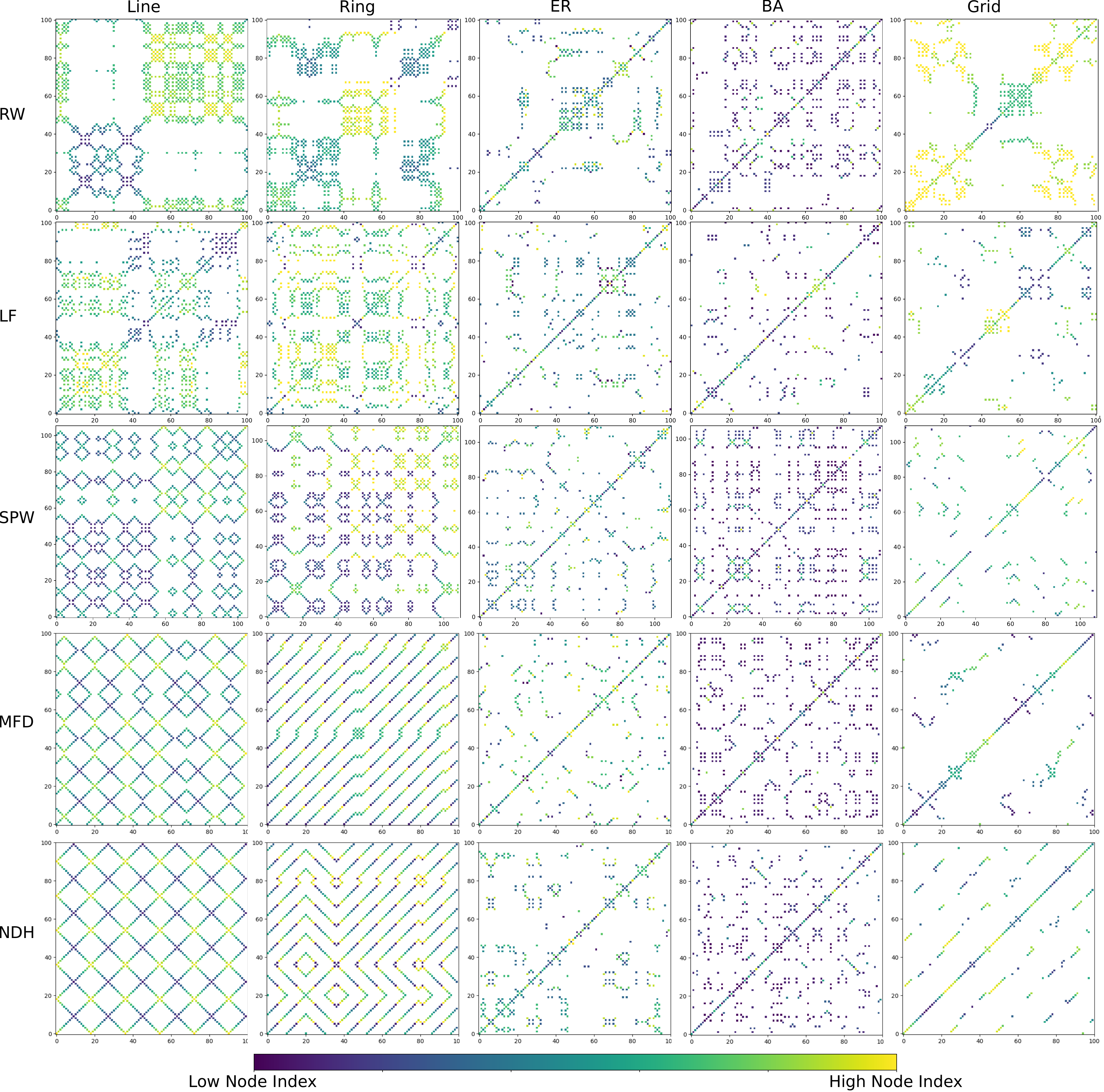}
    \caption{Recurrence plots for all combinations of networks and path dynamics (for one path segment of 100 steps each). White areas indicate no recurrence, dark purple indicates recurrence of low index nodes, yellow indicates recurrence of nodes with a high node index.}
    \label{fig:3}
\end{figure*}

\section{Recurrence plots show effects of networks and dynamics}
Node-weighted recurrence plots for combinations of networks and dynamics are shown in Fig.~\ref{fig:3}. The recurrence plots are shown for one path of length 100. Upon visual examination, a few characteristics are noticeable.

The recurrence plots of the random walk paths (RW) are shown in the top row of Fig.~\ref{fig:3}. They generally show the greatest variance with large white areas interspersed with densely colored areas. They also generally stay longer in each region of the network. These are commonly known features of random walks \cite{marwan2007recurrence,garcia2014,ramdani2016,ramdani2018}.

The Levy flights (c.f. second row) produce similar recurrence plots as the random walks, but with softer transitions and more mixing caused by the jumps. This shows up as smaller white areas, shorter diagonals, and shorter anti-diagonals.

The recurrence plots of the shortest path walks (SPW), shown in the third row, are more homogeneous and feature more diagonals and anti-diagonals of varying lengths, especially in the line and ring network. In ER and BA networks, diagonals and anti-diagonals are less prominent, but the plots appear a little more mixed than the previous dynamics, covering the network more evenly. On the grid diagonals of lengths between 5 and 10 are common.

The dynamics of the two bottom rows show significantly longer diagonals, especially on the ring and line networks, which topologically support periodic motions. On the grid, true diagonals are visible only in the NDH case with a distance of around 25, which is approximately equal to the outer border of the grid. The MFD algorithm, on the other hand, appears to show some approximate diagonals with a period around the network size. suggesting that the route covers the network entirely before starting again. Note, however, that for NDH and MFD, all 100 routes are constructed simultaneously, with some units arranging into periodicity, while others remain less structured.

There are also striking commonalities and differences between the different network topologies.

\begin{table*}[htp]
\caption{Average values for the five recurrence measurements from 100 realizations on each of the five networks with each of the five different walk types.}
\begin{ruledtabular}
\begin{tabular}{c|c|ccccc}
Network & Walk & $L$ & $RR$ & $W$ & $SW$ & $ AL$ \\
\hline
\multirow{5}{4em}{Line} & RW & $0.00162 \pm 1e-09$ & $0.10711 \pm 1e-05$ & $0.00415 \pm 2e-08$ & $1.91794 \pm 0.002$ & $0.00163 \pm 1e-09$ \\
& LF & $0.00122 \pm 1e-10$ & $0.10836 \pm 5e-06$ & $0.0041 \pm 5e-09$ & $2.82437 \pm 0.0005$ & $0.00122 \pm 1e-10$ \\
& SPW & $0.00246 \pm 1e-08$ & $0.12101 \pm 5e-06$ & $0.00362 \pm 5e-09$ & $2.26608 \pm 0.001$ & $0.00247 \pm 1e-08$ \\
& MFD & $0.01492 \pm 0.001$ & $0.13379 \pm 0.0005$ & $0.03342 \pm 0.01$ & $2.15462 \pm 0.05$ & $0.03193 \pm 0.01$ \\
& NDH & $0.20937 \pm 0.05$ & $0.10535 \pm 5e-07$ & $0.02241 \pm 5e-08$ & $2.28764 \pm 0.0005$ & $0.2179 \pm 0.05$ \\
\hline
\multirow{5}{4em}{Ring} & RW & $0.0015 \pm 2e-11$ & $0.1012 \pm 1e-06$ & $0.00442 \pm 5e-09$ & $2.15257 \pm 0.001$ & $0.00151 \pm 1e-11$ \\
& LF & $0.00119 \pm 5e-11$ & $0.10059 \pm 2e-07$ & $0.00445 \pm 5e-10$ & $2.96538 \pm 0.0005$ & $0.0012 \pm 5e-11$ \\
& SPW & $0.00205 \pm 5e-09$ & $0.10045 \pm 2e-07$ & $0.00446 \pm 5e-10$ & $2.4724 \pm 0.0002$ & $0.00202 \pm 2e-09$ \\
& MFD & $0.01441 \pm 0.001$ & $0.10878 \pm 0.001$ & $0.06473 \pm 0.02$ & $1.97929 \pm 0.1$ & $0.01249 \pm 0.002$ \\
& NDH & $0.18147 \pm 0.05$ & $0.10017 \pm 5e-08$ & $0.03098 \pm 2e-08$ & $0.84109 \pm 0.2$ & $0.06434 \pm 0.005$ \\
\hline
\multirow{5}{4em}{ER} & RW & $0.00118 \pm 2e-10$ & $0.02519 \pm 5e-07$ & $0.01899 \pm 5e-07$ & $4.02505 \pm 0.002$ & $0.0012 \pm 1e-10$ \\
& LF & $0.00108 \pm 5e-11$ & $0.02416 \pm 5e-07$ & $0.0198 \pm 2e-07$ & $4.43067 \pm 0.001$ & $0.0011 \pm 5e-11$ \\
& SPW & $0.00131 \pm 5e-10$ & $0.02802 \pm 5e-07$ & $0.01706 \pm 2e-07$ & $4.18948 \pm 0.002$ & $0.0013 \pm 5e-10$ \\
& MFD & $0.00163 \pm 5e-08$ & $0.02863 \pm 1e-06$ & $0.02197 \pm 5e-06$ & $4.16131 \pm 0.005$ & $0.00166 \pm 5e-08$ \\
& NDH & $0.00317 \pm 2e-08$ & $0.02927 \pm 2e-06$ & $0.03907 \pm 5e-06$ & $4.03754 \pm 0.005$ & $0.00325 \pm 2e-08$ \\
\hline
\multirow{5}{4em}{BA} & RW & $0.00118 \pm 5e-11$ & $0.03472 \pm 5e-06$ & $0.01373 \pm 5e-07$ & $3.68532 \pm 0.005$ & $0.0012 \pm 1e-10$ \\
& LF & $0.00108 \pm 5e-11$ & $0.02951 \pm 1e-06$ & $0.01619 \pm 5e-07$ & $4.20135 \pm 0.002$ & $0.00109 \pm 5e-11$ \\
& SPW & $0.0012 \pm 1e-10$ & $0.0541 \pm 5e-06$ & $0.00867 \pm 1e-07$ & $3.31998 \pm 0.005$ & $0.00119 \pm 1e-10$ \\
& MFD & $0.00196 \pm 5e-08$ & $0.04492 \pm 5e-06$ & $0.01711 \pm 5e-06$ & $3.52286 \pm 0.005$ & $0.00199 \pm 5e-08$ \\
& NDH & $0.00395 \pm 2e-08$ & $0.04805 \pm 1e-05$ & $0.03074 \pm 5e-06$ & $3.39816 \pm 0.01$ & $0.00405 \pm 2e-08$ \\
\hline
\multirow{5}{4em}{Grid} & RW & $0.00121 \pm 1e-10$ & $0.02273 \pm 1e-06$ & $0.02107 \pm 5e-07$ & $3.53532 \pm 0.005$ & $0.00123 \pm 1e-10$ \\
& LF & $0.00109 \pm 2e-11$ & $0.02226 \pm 2e-07$ & $0.0215 \pm 2e-07$ & $4.42 \pm 0.001$ & $0.00111 \pm 5e-11$ \\
& SPW & $0.00157 \pm 2e-09$ & $0.02211 \pm 2e-07$ & $0.02164 \pm 2e-07$ & $4.10701 \pm 0.0005$ & $0.00158 \pm 2e-09$ \\
& MFD & $0.00297 \pm 2e-07$ & $0.02215 \pm 5e-07$ & $0.0501 \pm 5e-05$ & $3.92875 \pm 0.002$ & $0.00305 \pm 2e-07$ \\
& NDH & $0.00678 \pm 5e-07$ & $0.02196 \pm 2e-07$ & $0.08623 \pm 5e-06$ & $3.91223 \pm 0.01$ & $0.00668 \pm 2e-07$ \\
\end{tabular}
\end{ruledtabular}
\label{tab.1}
\end{table*}
The first column of Fig.\ref{fig:3} shows recurrence plots of all dynamics on a line. As the line structurally only allows two directions, diagonals of sequences visiting the same nodes in the same order on this topology must be interspersed with the backtracking route of the same nodes in reverse order, which shows up as perpendicular lines in the recurrence plot. This pattern is best visible in the MFD and NDH recurrence plots, which show criss-cross patterns with few or no disturbances. These patterns are also visible in the SPW but barely or not at all for RW and LF.

The second column shows dynamics on the ring network. MFD and NDH produce the strongest patterns, with less backtracking than on the line. Some routes occasionally change direction, but diagonals dominate the plots.

The third column shows dynamics on Erdős-Renyi networks. As the network is larger than the line and ring networks, the overall recurrence rate is lower. Furthermore, due to the short average shortest path lengths, all dynamics appear highly mixed, covering the network more evenly, with smaller unbroken white areas.

The fourth column shows the recurrence in the dynamics on Barabasi-Albert networks. With its yet shorter average shortest path lengths, the recurrent points are very well distributed across the plot. For all dynamics, the dominating color is dark blue, indicating low node indices. The low indices mark the oldest nodes of the network, which are most likely the most central ones with the largest degrees.

The fifth column shows the recurrence plots for dynamics on the grid. This topology has a larger average shortest path length and no dead ends. Grid networks support periodicity, which can be seen as visible diagonals in SPW, MFD, and NDH dynamics.

\section{Recurrence measures can classify paths and networks}

To evaluate more than a small slice of a single trajectory, we compute a range of recurrence measures and search for clusters. The measures are selected to correspond to the heuristic findings we observe in Fig.~\ref{fig:3} and include several common RQA measures along with a new measure.

The recurrence rate \textit{RR} is defined as 
\begin{equation}
    RR=\frac{1}{N^2}\sum_{i,j=1}^{N} V_{i,j}/u_i,
    \label{eq.RR}
\end{equation}
where $N$ is the length of the trajectory, $V_{i,j}$ is the value-weighted recurrence matrix and $u_i$ is the value of the time series at time $i$. Dividing by $u_i$ removes the node index, treating all recurrences equally.

The mean diagonal line length \textit{L} is defined as
\begin{equation}
    L=\frac{\sum_{l_{\min}}^{N}l P_l(l)}{\sum_{l=1}^{N} P_l(l)},    \label{eq.L}
\end{equation}

where $P_l (l )$ is the length distribution of the diagonal lines of any non-zero value and we select $l_{min}=2$ as the minimal diagonal line length. In determining the diagonal line lengths, we again ignore node index or recurrence plot color.

The white vertical line length \textit{W}, also known as recurrence time \cite{schinkel2008selection} is defined as 
\begin{equation}
    W=\frac{\sum_{w=1}^{N} w P_w(w)}{\sum_{w=1}^{N} P_w(w)},
    \label{eq.W}
\end{equation}
where $P_w(w)$ is the distribution of white vertical line lengths.

The white vertical entropy $S_W$ is defined as 
\begin{equation}
    SW=-\sum_{w=1}^{N}  \frac{ P_w(w)}{\sum_{w=1}^{N} P_w(w)} \ln  (\frac{ P_w(w)}{\sum_{w=1}^{N} P_w(w)}),
    \label{eq.SW}
\end{equation}

\begin{figure}[h]
    \centering
    \includegraphics[width=\columnwidth]{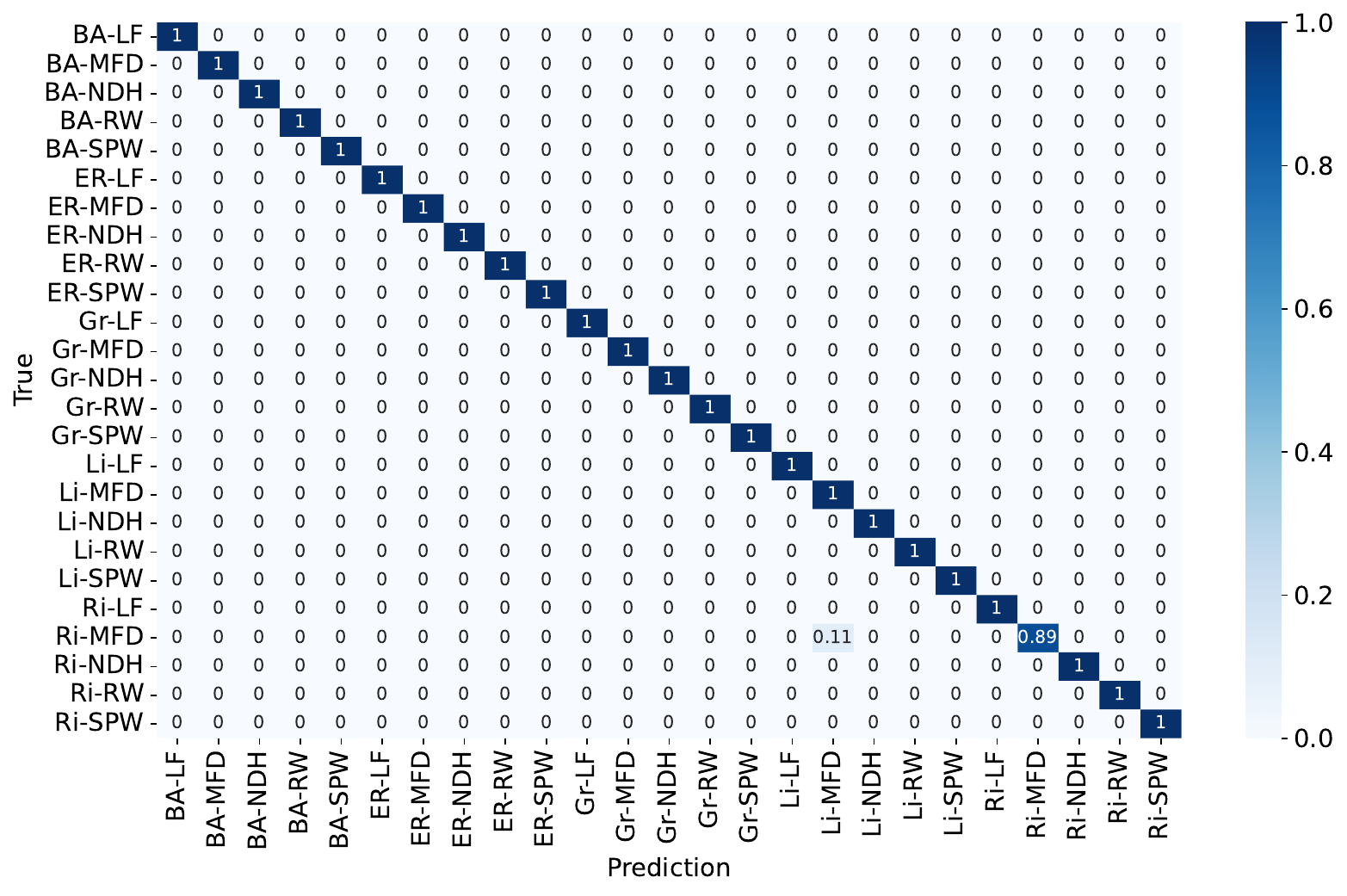}
    \caption{The five recurrence measures and the route lengths from Tab.~\ref{tab.1} are sufficient to classify the routes with a random forest classifier.}
    \label{fig:4}
\end{figure}

\begin{figure*}[htb]
    \centering
    \includegraphics[width=\textwidth]{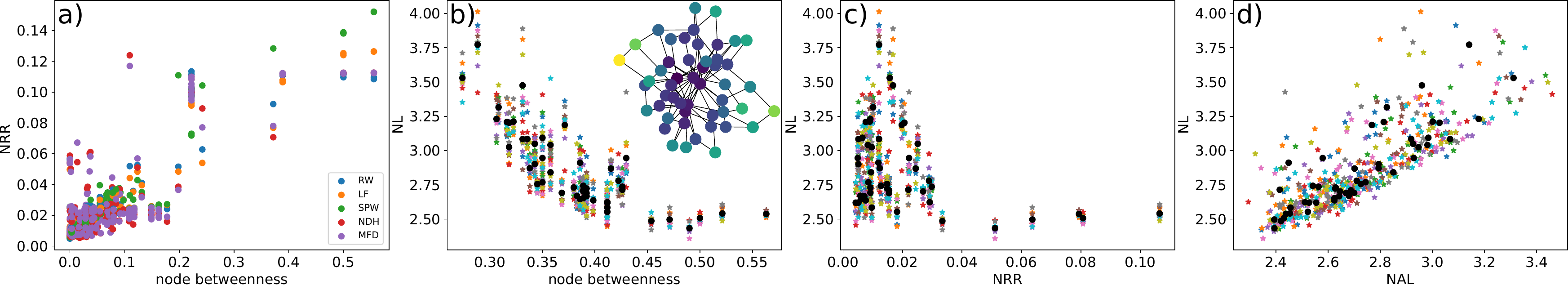}
    \caption{Node-based recurrence measures vary with centrality. a) Node recurrence rate increases with node betweenness. Across all networks, central nodes are more frequently visited by all dynamics. For the routing algorithm, this effect is weakest, with NDH and MFD showing the most mixed behaviour out of all dynamics. b) Node mean diagonal line lengths for 10 realizations (and their average in black) of SPW on BA network plotted over node betweenness show a decreasing behaviour. Central nodes are part of shorter diagonals than the outer nodes. The averages are shown as node colors on the graph in the inset. c) Node mean diagonal line lengths over node recurrence rate shows that less recurrent nodes are part of longer diagonals. d) node mean anti-diagonal line length and node mean diagonal line length are positively correlated, but diagonals are slightly longer on average than anti-diagonals.}
    \label{fig:5}
\end{figure*}

To describe the common appearance of perpendicular diagonals in the plots of Fig.~\ref{fig:3}, we define the average anti-diagonal line length $AL$ as

\begin{equation}
    AL=\frac{\sum_{a_{\min}}^{N}a P_a(a)}{\sum_{a=1}^{N}P_a(a)},
    \label{eq.aL}
\end{equation}
which is defined analogously to the average diagonal line lengths but based on the distribution $P_a(a)$ of anti-diagonal line segments $a$ of non-zero values in the recurrence plot, again selecting $a_{min}=2$.

This pattern indicates time reversal in the time series, or backtracking of the walker in our case. While reversals are relatively common for trajectories on sparse networks, they are less widespread in spatially embedded dynamical systems.

The average for each measure across all eligible realizations of all combinations, (100 simulations were performed but 58 trajectories from MDH dispatcher on the ring were too short for consideration), is shown in Tab.~\ref{tab.1}. Closely examining the recurrence measure values and comparing all pairs of lines, we find that they are distinguishable. With five recurrence measures and the route length, the 25 combinations can also be distinguished well by training a random forest classifier with 300 trees. The 100 trajectories for each combination are split 80:20 into training and test sets. For each trajectory we calculate the six measures, resulting in 2442 input vectors (58 trajectories from the MFD tests are too short to compute meaningful recurrence measures). The average cross-validation score is 99.6\% with a standard deviation of 0.5\%. Fig.~\ref{fig:4} shows the confusion matrix of the 25 combinations, revealing that only the MFD routes on the ring are misclassified in 11\% of the routes. The misclassification is likely related to the lower number of realizations in this case. The MFD and NDH dispatchers were simulated for a fleet of 100 walkers. However, the MFD optimization reduces the driven distance by assigning requests to fewer routes, leaving some walkers standing still. In the route analysis, we have chosen to neglect the empty routes. For all other combinations, the classifier has learned to correctly predict them. Route classification on larger networks of $N\approx 1000$ was done in the supplemental material. While accuracy is still good, there it only reaches an average of 94\%.

\section{node-based diagonal counts}

The above selection of recurrence measures only uses the unweighted recurrence matrix. Here, we introduce three new measures that make use of the node information encoded in $V_{i,j}$.

We define the \textit{node recurrence rate} $NRR(n)$ of node $n$ as:
\begin{equation}
    NRR(n)=\frac{1}{N^2}\sum_{i,j=1}^{N} \delta (V_{i,j}-n),
    \label{eq.NRR}
\end{equation}
where the $\delta$-function is 1 if $n=V_{i,j}$ and 0 otherwise. Thus, NRR counts the recurrence rate of each node separately.

We further define the \textit{average node diagonal line length} $NL(n)$ of node $n$ as:
\begin{equation}
    NL(n)=\frac{\sum_{l_{\min}}^{N}l P_l(n,l)}{\sum_{l=1}^{N} P_l(n,l)},
    \label{eq.NL}
\end{equation}
where $P_l(n,l)$ is the length distribution of diagonals including node $n$ and $l_{min}=2$.

Similarly, the \textit{average node anti-diagonal line length} $NAL(n)$ of node $n$ is defined as:
\begin{equation}
    NAL(n)=\frac{\sum_{a_{\min}}^{N}a P_a(n,a)}{\sum_{a=1}^{N} P_a(n,a)},
    \label{eq.NL}
\end{equation}
where $P_a(n,a)$ is the length distribution of anti-diagonals including node $n$ and $a_{min}=2$.

These measures are shown in various combinations in Fig.~\ref{fig:5}. The NRR is plotted over node betweenness in Fig.~\ref{fig:5} for trajectory lengths of 2000, averaged over 100 realizations for all 25 combinations. The node recurrence rate increases with node centrality across all networks and dynamics, indicating that paths use the most central nodes more frequently than the less central ones. This effect is weakest for the two routing algorithms, which show a wider distribution of NRR with betweenness.

In contrast, the average node anti-diagonal line length is negatively correlated with node betweenness, shown in Fig.~\ref{fig:5}b for SPW dynamics on a Barabasi-Albert graph. Although peripheral nodes incur fewer visits to the paths, those visits have a higher likelihood of being part of a longer repetitive pattern. This suggests that reaching peripheral nodes necessitates passing through one of a few sequences of nodes before and/or after.

Together, the findings in Fig.~\ref{fig:5}a) and b) imply that node recurrence rate and node mean diagonal line length are anti-correlated, again for SPW dynamics on a Barabasi-Albert graph (Fig.~\ref{fig:5}c). The node mean diagonal line length and node mean anti-diagonal line length are correlated for the same setting (Fig.~\ref{fig:5}d).

 \section{Discussion and conclusion}

In this work, we explore the use of node-weighted recurrence plots and RQA for five distinct types of walks across five networks. This analysis counts returns of a path to the same node and encodes it as a weight or color in the recurrence plot. Although chromatic/symbolic recurrence analysis is conceptually similar in its use of a weight or color to encode additional information, here the additional information is spatial or topological, and thus partially dependent on the temporal dimension via path lengths. As a result, the patterns observed in the node-weighted recurrence plots are structurally distinct from the patterns observed in the chromatic/symbolic recurrence plots, which are dominated by rectangular structures and vertical/horizontal lines \cite{cox2016chromatic, caballero2018symbolic}. Node-weighted recurrence plots show a broad spectrum of patterns, including diagonal lines and anti-diagonal lines, but few or no vertical/horizontal lines. 

While diagonal lines are common in recurrence plots from dynamical systems, anti-diagonals result from time-reversal, which is common for paths on small networks with dead ends, but less common for spatially embedded dynamical systems. We find that for most dynamics and networks, diagonals and anti-diagonals are approximately equal in length. Analyzing 100 realizations of each combination of dynamics and networks using standard RQA measures (Eq.~\ref{eq.RR} to Eq.~\ref{eq.aL}), we have shown that the RQA measures are sufficient to distinguish between the trajectories through direct comparison to the means of each combination or a random forest classifier.

All results discussed so far were based on existing color-agnostic recurrence measures. To account for the additional information encoded in the recurrence matrix weights, we have introduced the node recurrence rate, the average diagonal line length, and the average anti-diagonal line length. These measures only count the recurrent elements that match a given node. This reveals that the most central nodes have a higher recurrence rate, but relatively lower diagonal and anti-diagonal line lengths. This indicates that peripheral nodes are less often visited, but if they are, it is more frequently along the same paths. In a human mobility setting, this would indicate a higher predictability.

Our results show that node-weighted recurrence analysis improves our understanding of differences across dynamics and networks, as well as within nodes in the same network. These insights could be crucial for improving the routing for pooled ride-hailing services. Further research is needed to connect our findings to routing research to avoid jamming and to extend the analysis to track higher-dimensional processes such as disease spreading. Additionally, further research is needed to verify the analysis for larger or inhomogeneous networks.

\begin{acknowledgments}
\textit{Acknowledgments.} Alexander Schmaus acknowledges support from the German Federal Environmental Foundation (Deutsche Bundesstiftung Umwelt).

\textit{Data availability.} The data that support the findings of this study will be made publicly available in the Zenodo.
\end{acknowledgments}

\bibliographystyle{unsrt}
\bibliography{pathrec,rp}

\end{document}